# Spatial-offset pump-probe imaging of nonradiative dynamics at optical resolution

**Authors**

Guo Chen[1†], Yuhao Yuan[1†], Hongli Ni[1], Guangrui Ding[1], Mingsheng Li[1], Yifan Zhu[1], Deming Li[1], Hongru Zeng[1], Hongjian He[1], Zhongyue Guo[2], Ji-Xin Cheng[1,2]*, Chen Yang[1,3]*

**Affiliations**

[1] Department of Electrical and Computer Engineering, Boston University, Boston, MA 02215, USA

[2] Department of Biomedical Engineering, Boston University, Boston, MA 02215, USA.

[3] Department of Chemistry, Boston University, Boston, MA 02215, USA.

* Corresponding authors: jxcheng@bu.edu; cheyang@bu.edu
† These authors contributed equally to this work

**Abstract**

Nonradiative photothermal (PT) and photoacoustic (PA) processes have found widespread applications in imaging, stimulation, and therapy. Mapping the generation and propagation of PA and PT waves with resolution is important to elucidate how these fields interact with biological systems. To this end, we introduce spatial offset pump-probe imaging (SOPPI). By spatially offsetting the pump beam and the probe beam, SOPPI can image simultaneously PA and PT wave propagation with nanosecond temporal resolution, micrometer spatial resolution, 65 MHz detection bandwidth, and a sensitivity of 9.9 Pa noise equivalent pressure. We first map the PA and PT evolution from a fiber emitter, and how the wave interacting with a mouse skull and brain slices. SOPPI imaging of PA waves from a tapered fiber with water as an absorber shows a wavelength-dependent generation, evanescent wave generated PA, and back-propagated acoustic Mach Cone. At last, a SOPPI-PACT is developed to reconstruct the pigment distribution inside a zebrafish larva with high precision and signal-to-noise ratio.

**Teaser**

Spatial offset pump-probe imaging (SOPPI) maps photothermal and photothermal fields at optical spatial resolution and nanosecond temporal resolution.

# MAIN TEXT

# Introduction

Non-radiative relaxation is a process of energy transfer without emitting photons when an excited state of a molecule or atom returns to a lower energy state. The energy released during the transition dissipates as heat transferring to other atoms or molecules in the surrounding environment. During the heat dissipation, the rising pressure caused by the rapid thermal expansion at given conditions could generate acoustic waves. For example, with nanosecond pulsed laser excitation, photoacoustic (PA) and photothermal (PT) effects occur simultaneously when thermal and stress confinements were met (*1*). Both the thermal and acoustic effects in nonradiative relaxation play important roles when interacting with biological systems and have been deployed in sensing (*2-6*), imaging (*7-12*), therapy (*13-16*), neural stimulation (*17-24*), and more (*25*). Visualization of PA and PT processes opens opportunities to understand the mechanism of these processes and their functions in intervening biological systems.

Despite the significance, direct and simultaneous visualization of PA and PT has not been achieved. Temperature and pressure detection have not been integrated through a single platform. Traditionally, for PA detection, an ultrasound transducer or hydrophone is used. They have limited bandwidths and acceptance angles (*26, 27*), which limits their applications in measuring PA signals with a broad frequency range and wide spatial distribution. Newly developed optical cavity-based PA measurement has improved the bandwidth and acceptance angles (*28-30*). However, like the transducer and hydrophone, such devices cannot be used to measure signals within solid samples or in situ. For PT detection, thermographic cameras have been used widely (*31*). However, the spatial and temporal resolution of thermographic cameras are limited by the detected wavelength of infrared light at a few microns and by the imaging speed of a few tens of frames per second.

Optical pump-probe techniques have been developed for detecting shockwave (*32*), PA (*33*), and PT (*34*) fields. Refractive index change induced by shockwave, PA or PT with a pulsed pump beam could be observed by introducing a probe beam. Through co-localization and scanning of the two beams, the absorbers can be mapped out. With such method, Zemp and coworkers reported photoacoustic remote sensing of blood vessels under the skin (*2*). More recently, Ni et al. used short-wave IR to excite the localized PT signal from lipids in biological samples and a probe beam to detect the refractive index change induced (*35*). While point scanning pump-probe imaging measures the local refractive index change at the absorbers, the environmental change caused by the PA and PT propagation in the surrounding medium is concealed. Beyond point scanning pump-probe imaging, wide field time-resolved interferometric pump-probe imaging is commonly used to measure wave propagation with a higher frame rate (*36*). Lately, streak cameras have been used for single-shot shockwave imaging (*37, 38*). Despite these advances, the limited memory of the camera hinders the recording of the full PA and PT lifetimes ranging from nanoseconds to micro- or even milliseconds. A method that can map the spatial and temporal evolution of both PA and PT fields is lacking.

In this work, we introduce spatial offset pump-probe imaging (SOPPI) to visualize the nonradiative dynamics with nanosecond temporal resolution. Distinct from conventional point scanning co-localized pump-probe imaging, SOPPI spatially separates the pump and probe beams in order to detect the refractive index change both at the absorbers and in the surrounding where the fields propagate through. By fast digitization of the probe photons at 180 MHz, SOPPI is able to map the spatial and temporal evolution of PT and PA fields. Our pump-probe approach offers a spatial resolution of 6 microns when detecting the pressure, 10 times better the typical hydrophone detection. Taking advantage of optical detection, SOPPI realizes pascal-level sensitivity by detecting changes in the refractive index smaller than 0.01 induced by non-radiative relaxation (*2*). The technique also offers a broad detection bandwidth, covering ultrafast PA signals in the tens of MHz and slow PT signals that can persist for tens of microseconds. As applications, we demonstrate the first-time observation of near-field physical processes, including acousto-thermal effects and evanescent wave-generated PA waves. Owing to its highly sensitive and quantitative detection of PA signal, SOPPI serves as a programmable virtual transducer array with an optical density, and can be coupled with photoacoustic computed tomography, termed as SOPPI-PACT, for mapping of optical absorbers inside a small animal. Overall, SOPPI enables direct and simultaneous visualization of the full evolution of PA and PT waves and their interactions with biological tissue, offering a platform technology for the mechanistic study of PT/PA generation and rational design of PT/PA-based medical devices.

**Results**

**SOPPI captures PA and PT signals simultaneously with optical spatial resolution and nanosecond temporal resolution**

To simultaneously visualize the PA and PT signals, we spatially separated the pump and probe foci. As shown in **Figure 1a**, we have the probe beam fixed and scan the fiber emitter mounted on a 2D scanning state to map out the subtle change of refractive index induced by PA and PT effects and their evolution in the environment. The scanning step size is adjustable, ranging from 0.1 μm to 40 μm. This flexibility allows the total scanning area to vary from micrometer to centimeter scales.

To generate PA and PT fields, a 1064 nm, 5 ns pulsed laser was used as the pump. The pump laser was delivered through a multimode fiber to an absorber. A 1310 nm continuous wave laser was used to probe the generated fields. The probe beam was delivered through a 10x water-dipping objective and was collected by a 10x air objective. The collected probe beam was then detected by an amplified InGaAs photodiode with a 1310 nm bandpass filter and converted to electrical signals by a fast digitization card. The detected intensity is captured in the direct current (DC) channel. Details of the components can be seen in the Method section. When the generated PA/PT signal interacts with the probe beam, the electrical signals will also be modulated and detected in the alternating current (AC) channel. The acquisition of signals is synchronized with the pump laser and the spatial scanning of the motorized translation stage. As a demonstration of spatial offset imaging of PA and PT waves, we used a fiber emitter (FE) to efficiently generate both PA and PT fields. The FE was made of a multimode optical fiber (FT200EMT, Thorlabs), with a layer of candle soot and a second layer of polydimethylsiloxane (PDMS) coated on the tip of the fiber. The strong absorption of candle soot and the large expansion coefficient of PDMS make this device an efficient PA emitter and PT generator as well (*17*).

A representative unprocessed signal generated by the fiber emitter in water was displayed as a function of time (**Figure 1b**). Two components, including a high-frequency bipolar PA signal (**Figure 1b**, green dash box) and a low-frequency slowly decaying PT signal (**Figure 1b**, red dash box), were observed. Since the central frequency of PA and PT signals are well separated (PA at MHz and PT < kHz), a high pass filter with a cutoff frequency at 500 kHz can effectively distinguish both signals (**Figure 1c**). A Hilbert transform was applied to show the intensity of the PA signal (**Figure 1c**, dashed blue line).

By measuring a 3 µm poly(methyl methacrylate) (PMMA) bead as the absorber (**Figure 1d**), we showed that the spatial resolution of the system is 6.1 µm, which matches the theoretical value calculated with 1310 nm wavelength of the probe laser and an effective numerical aperture of 0.13. The temporal resolution of 5.56 ns of the system was determined by the 180 MHz sampling speed of the digitizer. We also characterized the pressure sensitivity of our SOPPI system by measuring the PA signals and comparing them to the results measured by a commercial hydrophone (HGL-0085, Onda) (**Figure 1e**). The distance between the FE tip and the hydrophone/probe beam was precisely controlled to be 300 microns by maintaining the delay between the trigger of the pump beam and the signal. The SOPPI signal showed a broader detection bandwidth (FWHM = 51 MHz) (Yellow, **Figure 1e**), with a central frequency at 41 MHz while the hydrophone result showed a frequency spectrum centered at 20 MHz with an FWHM = 38 MHz (Orange, **Figure 1e**). We calibrated the SOPPI signal intensity with the hydrophone measured PA pressure using different pump beam pulse energy from 5 nJ to 15 µJ (**Figure 1f**). Both the hydrophone and SOPPI results showed a linear relationship between the signals and input energy ($R^2$=0.9982 for hydrophone measurement, $R^2$=0.9988 for SOPPI) (**Figure 1f**). The lower boundary of the detection limit of the hydrophone was reported to be approximately 1 kPa (*39*). SOPPI shows a noise equivalent pressure (NEP) of 9.8 Pa (**Figure S1**), which is two orders better than the hydrophone, and one order better than piezo based transducers (*40*). Compared with other optical detection methods, SOPPI's NEP is also better than most interferometry-based detection methods and comparable to micro-ring resonators (*41*). In summary, SOPPI offers high sensitivity at the order of ten pascals, high spatial resolution at the optical diffraction limit, and high temporal resolution at the nanosecond level in detecting photoacoustic signals and allows observation of subtle changes of pressure/temperature that are unreachable by previous methods.

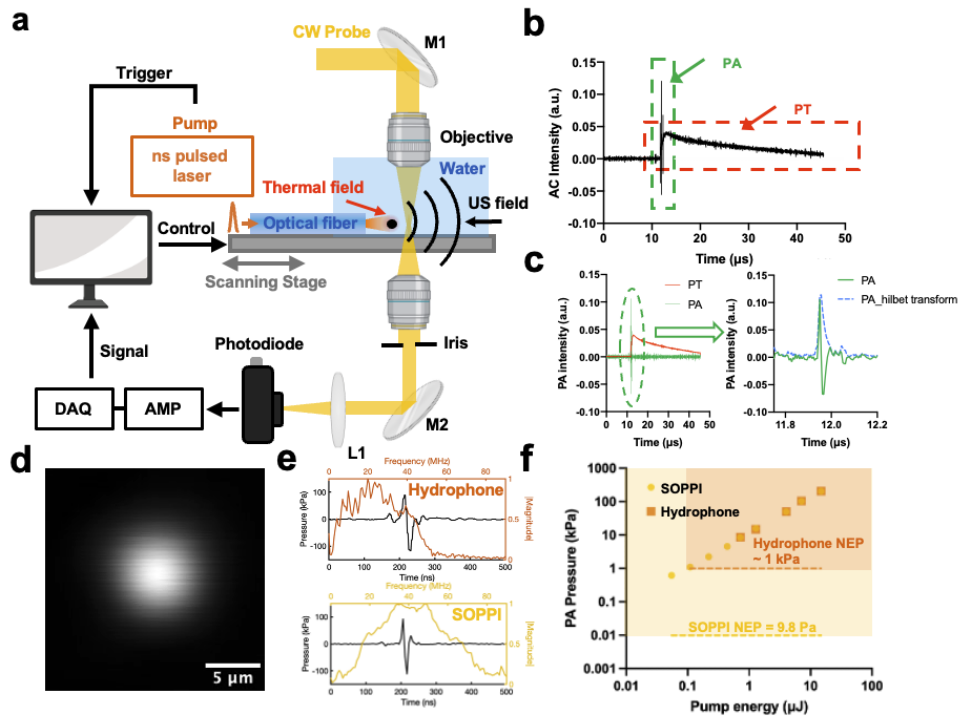

**Figure 1. SOPPI is a spatially offset pump-probe method to measure the PA and PT fields generated by a fiber emitter. a.** Schematic of the SOPPI setup. **b.** Representative AC

signal detected by SOPPI. **c.** Decomposition of the raw signal. Red: photothermal signal. Green: photoacoustic signal. Blue: photoacoustic signal after Hilbert transform. **d.** Spatial resolution characterization of SOPPI using a 3 μm PMMA bead. **e.** PA signals measured by hydrophone (orange) and SOPPI (yellow). **f.** Peak-to-peak pressure measured by hydrophone and SOPPI and plotted as a function of pump energy.

**SOPPI reveals complex evolution of PT and PA fields in a non-radiative relaxation.**

By spatially offsetting the pump and probe beam, SOPPI allows simultaneous visualization of PT and PA fields. We measured the PA/PT dynamics generated from the FE by spatially scanning the FE sample and the adjacent environment with the probe laser while the pump laser repeatedly illuminated the FE sample. Through a single scanning, the PA and PT fields were mapped out simultaneously by SOPPI. **Figure 2a** displays an overlaid image of PA (green) and PT (red) signals at every 100 ns from 0 ns to 1400 ns. **Figure 2a** and **Video S1** show that while the PA signal propagated along the x-axis (green), the PT signal was localized (red).

With a broad detection bandwidth, SOPPI shows that PA waves with different frequencies behave differently when propagating. Noting that the high-frequency ultrasound propagated more directional, and the low-frequency ultrasound propagated more omnidirectional, we decomposed the raw signal into a high-frequency component and a low-frequency component by applying a 10 MHz high/low pass digital filter. The representative wavefront of each component at 400 ns (**Figure 2b** and **2c**), the corresponding PA signal and the frequency spectrum (**Figure 2d** and **2e**) were shown. We plotted the PA signal amplitude as a function of the location of both the high-frequency PA component (**Figure 2b**, bottom) and the low-frequency component (**Figure 2c**, bottom). From this result, the wavelength of the high-frequency PA was measured to be 75 μm and the wavelength of the low-frequency PA was 150 μm, consistent with the frequency analysis shown in **Figure 2d** and **e**.

Interestingly, the two different PA components show distinct spatial distribution. With higher frequency and smaller wavelength, the high-frequency component generated a focus at 120 μm away from the fiber tip (**Figure 2f**, yellow). This is because the spherical shape of the PDMS coating formed an acoustic lens and focused the ultrasound. In contrast, the low-frequency component of the ultrasound (**Figure 2f**, blue) started to decay immediately after being emitted from the FE sample. This is because that the 200 μm diameter of the FE, which also acts as the aperture of the acoustic lens, made it hard to focus the low-frequency ultrasound with a wavelength of 150 μm. Since the minimum detection element size of a commercialized needle hydrophone is 45 μm, it is impossible for such a hydrophone to map out a precise spatial field distribution of ultrasound with a wavelength smaller than 90 μm (twice the element size based on the Nyquist sampling law). Together, SOPPI provides a unique and superior capability to visualize the spatial distribution of the ultrasound with optical resolution.

To map the temporal evolution at given locations and spatial distribution of the PT field, we recorded a time trace of 45 μs to observe both its rise and fall (**Figure 2h**). We applied the maximum intensity projection of the PT traces to show the PT distribution inside and outside the fiber (**Figure 2g**). The maximum intensity projection was normalized by the direct current (DC) signal to better quantify the relative temperature increases at different locations. A noticeable temperature increase was observed inside the fiber emitter (**Figure 2i**). The boundary between the fiber and water acted as a thermal insulator, localizing the heat due to water's low thermal conductivity.

As an optical detection method, SOPPI can access information inside the absorber as long as the probe laser can be detected. To demonstrate this, we use SOPPI to visualize the spatial and temporal evolution of PA and PT signals inside the fiber emitter (**Figure S2**). The PA and PT signals are initially generated inside the coating of the FE within the first 50 ns (**Figure S2a**). After that, the PA signal (green) propagated out of the FE and was partially reflected due to the acoustic impedance mismatch at the PDMS/water boundary. Importantly, energy converts from mechanical wave into heat occurred at this boundary. To quantify this acousto-thermal process, we plotted the raw signal traces at two locations: one at the FE boundary (**Figure S2 a,b**, yellow) and the other inside the FE (**Figure S2 a,c**, blue). Signals from both locations exhibited high-frequency components initially, followed by prolonged PT heating lasting more than 20 µs. The boundary region experienced an additional secondary heating process on top of the PT baseline within the first 3 µs. These results show that with the optical spatial resolution and nanosecond temporal resolution, SOPPI reveals rapid near-field energy conversion not accessible with traditional detectors such as transducers.

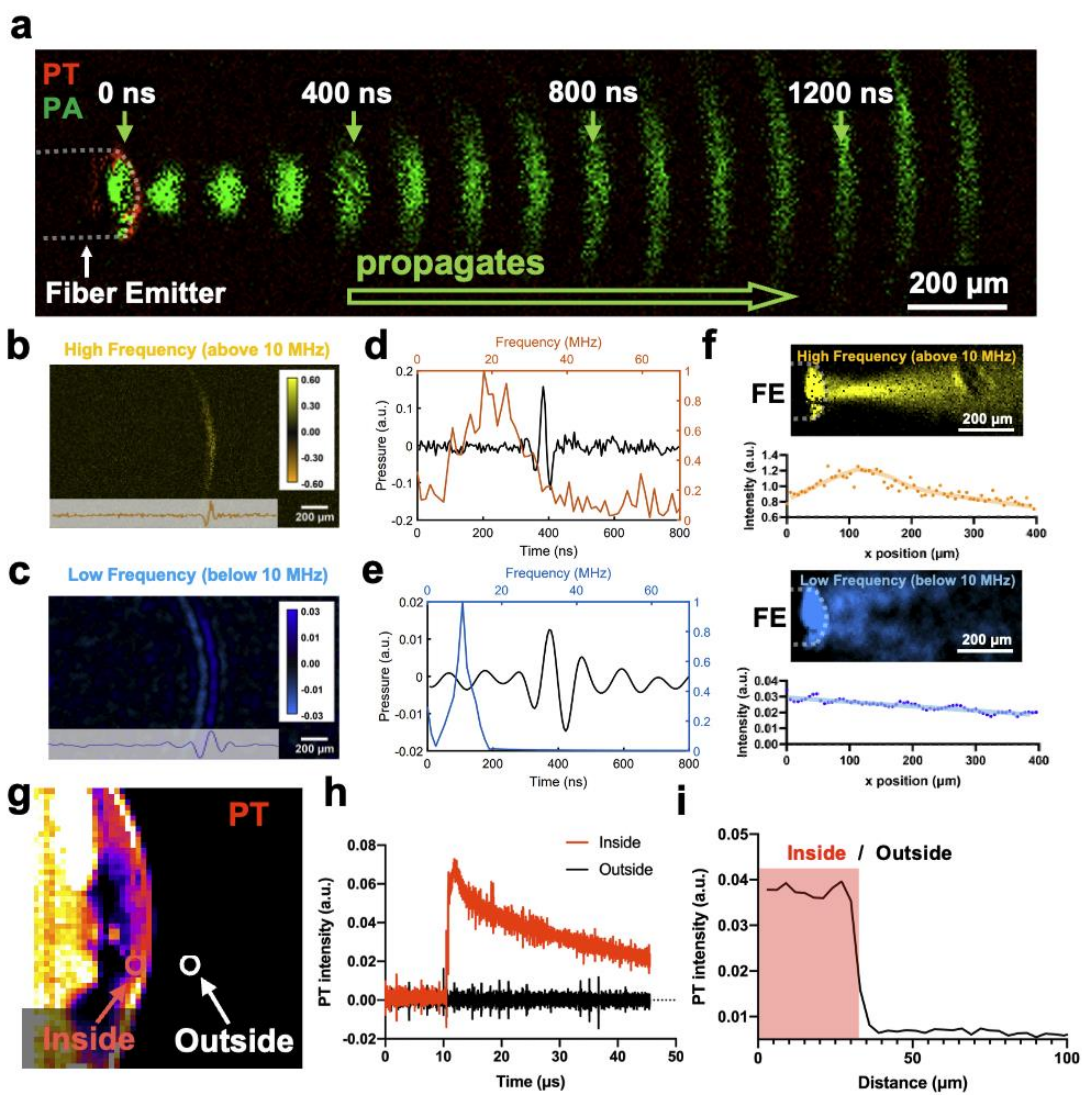

**Figure 2. SOPPI reveals the spatial evolution of PA and PT signal generated by a fiber emitter. a.** An overlaid image of PA (green) and PT (red) signals at every 100 ns from 0 ns to 1400 ns. Red: photothermal. Green: photoacoustic intensity. **b.** High-frequency component of the PA signal from **a**. Bottom: PA signal amplitude plotted as a function of location. **c.**

Low-frequency component of the PA signal from a. Bottom: PA signal amplitude plotted as a function of location **d.** Time trace and Fast Fourier transfer of the high-frequency component of the photoacoustic signal. **e.** Time trace and Fast Fourier transfer of the low-frequency component of the photoacoustic signal. **f.** Spatial distribution of high-frequency PA (orange) and low-frequency PA (blue). **g.** Spatial distribution of the normalized photothermal signal generated by a fiber emitter. **h.** Representative time trace of the photothermal signal inside and outside the FE. **i.** PT intensity plotted as a function of distance away from the FE.

**SOPPI maps the PT and PA fields generated by water via a tapered optical fiber.**

Water is a strong absorber in the short-wave infrared window and has been widely used for infrared PT neuromodulation (*42*). PA imaging of water content in skin at 1540 nm was recently demonstrated. The PA signal generated by skin water was used for ultrasound imaging (*43*). To better harness water as an endogenous PT and PA emitter, we acquired the nonradiative dynamics of water excited at wavelengths from 1200 nm to 2100 nm covering two major absorption bands of water. The tunable pump laser was delivered through a tapered optical fiber with a tip size smaller than 50 μm (**Figure 3a**). In this specific setup, the emitted laser generated an evanescent wave on the side of the fiber in the tapering region. Both the direct emission and the evanescent wave were absorbed by water, leading to the generation of PA and PT fields (**Figure 3b**). This complex interaction involved water absorption at different wavelengths, as well as various nonradiative emissions occurring in both the near and far fields, making it challenging to visualize using traditional methods.

By sweeping the wavelength of the pump laser, PA signals at the same location, 160 μm away from the tapered fiber tip, at two absorption peaks were recorded by SOPPI (**Figure 3c**). The generated PA amplitude as a function of wavelength exhibited two peaks at 1448 nm and 1928 nm (**Figure 3d**, red curve), consistent with the reported absorption spectrum of water (**Figure 3d**, black curve) (*44*). A frequency analysis was performed for wavelength-dependent water absorption (**Figure 3e**). The central frequency was indicated by the red solid line and the -6 dB bandwidth was shown by the white dashed line. The analysis revealed that at the absorption peak of 1928 nm, water generated a PA signal with a broader bandwidth compared to the signal at 1448 nm, which can be explained by the smaller absorption volume with higher optical attenuation at 1928 nm (*45*).

To visualize the spatial distribution of both PA and PT fields, we tuned the pump laser to 1928 nm, corresponding to the strongest absorption peak of water in the near-infrared (NIR) range. Propagation of ultrasound within 200 ns after the pump was recorded (**Video S2**). Frames at each 50 ns were displayed (**Figures 3f-j**), where the green and yellow represented the positive and negative amplitudes of PA, and the red represented the intensity of PT. A propagating PA field in water and a localized PT field at the boundary between the tapered fiber and water were observed.

Unlike the PA/PT fields generated by the FE described earlier, the tapered optical fiber generated fields not only at its tip but also along its edge, due to water absorption of evanescent waves in the tapering region of the fiber. When a multimode fiber is tapered, the core with a decreasing diameter no longer holds the higher-order modes, causing them to leak into the cladding layer of the fiber (*46*). Consequently, the cladding-water boundary acts as a total internal reflection interface within this waveguide. Stronger evanescent waves occurred on the water side due to total internal reflection at the water interface and were absorbed to produce PT (red along the fiber edge) and PA (**Figures 3j** and **3k**, blue boxes).

In addition to forward and lateral ultrasound emission, we also observed part of the signal propagating backward through the fiber. Given that the fiber is made of silica, a material with a sound speed three times higher than that of water, the back-propagated ultrasound formed a shock wave in water, creating a Mach cone with a triangular shape (**Figures 3j** and **3k**, orange boxes). Together, SOPPI captures the evolution of this complex physical process with its 10 Pa NEP sensitivity, optical spatial resolution, and nanosecond temporal resolution.

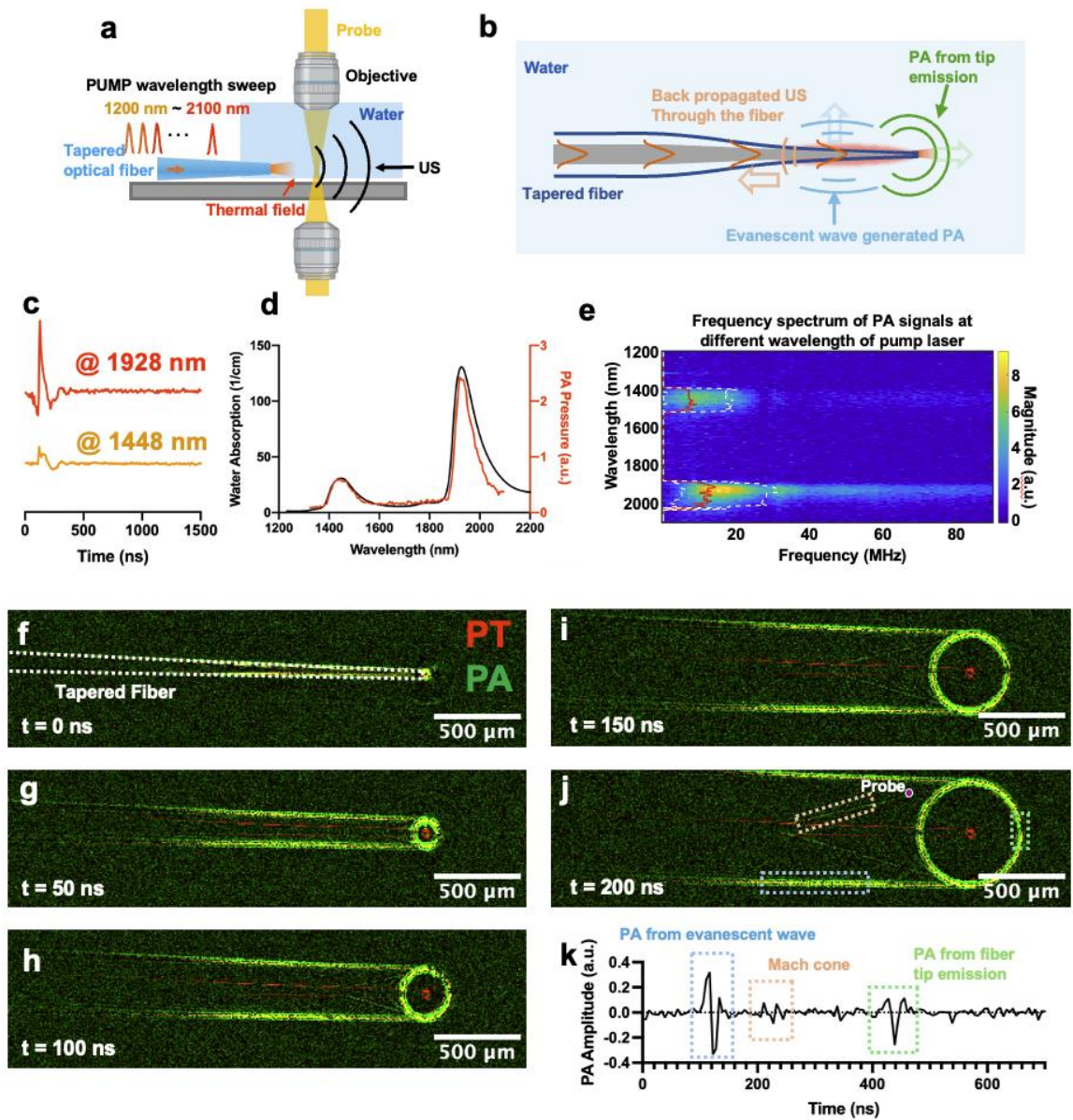

**Figure 3 SOPPI of PA/PT fields generated by water through a tapered optical fiber. a-b**. Schematic of the imaging setup. Near-infrared (NIR) nanosecond pulsed laser was coupled into a tapered optical fiber. Water served as the absorber as well as the acoustic coupling medium. **c.** PA signals generated from water at different wavelength of pump laser at a location 160 μm away from the tip. **d-e.** PA amplitude and frequency spectrum under different wavelength of pump laser in NIR. **f-j.** Spatial distribution of PA (green and yellow) and PT fields (red) at 0-200 ns after the laser was turned on. Dashed boxes: evanescent wave absorption (blue), Mach cone (orange), and tip of the tapered fiber (green) **k.** three PA waves from evanescent wave absorption (blue), Mach cone (orange), and tip of the tapered fiber (green) at 1920 nm pump. The signal was measured at the "probe" point labeled as purple.

**SOPPI maps the propagation of PT and PA fields in a scattering environment.**
Ultrasound has been widely used in brain imaging and stimulation. Understanding how ultrasound interacts with brain tissue and the skull is essential for designing and optimizing the technology for biomedical applications. Yet, visualizing wave propagation inside highly scattered biological samples including brain slices and skulls is not possible for physical devices like transducers or hydrophones. SOPPI offers this unique capability of imaging the interaction between ultrasound and tissues.

With a 1310 nm laser as the probe, we demonstrated SOPPI of ultrasound through brain slices with a thickness of 500 μm as an example of highly scattered objects (**Figure 4a-c**). The ultrasound went through the coronal plane of a mouse brain slice in phosphate-buffered saline (PBS). To quantitatively compare the ultrasound intensity before and after penetrating the brain, we first scanned the probe beam over the sample without the pump laser to obtain the DC image of the brain slice (**Figure 4a**). Next, with the pump laser activated, we captured the PA signal intensity map, which is the AC signal (**Figure 4b**). Since the AC signal amplitude can be significantly influenced by the DC intensity in complex biological samples, we normalized the AC mapping results using the DC image (**Figure 4c**). This approach enables a quantitative comparison of the PA intensity distribution outside and inside the brain slice. After the normalization, the incident and transmitted PA signal amplitude is plotted as a function of time in **Figure 4d**. By plotting the ultrasound amplitude as a function of distance to the acoustic source (**Figure 4e**), we did not observe significant decay and reflection of PA waves when propagating through the brain/PBS interface. This finding suggests that more than 99% of the ultrasound energy delivered into the brain tissue, which can be attributed to the low acoustic impedance mismatch between PBS and brain tissue. This result is consistent with the calculation results (See Supplementary Information). In ultrasound imaging and stimulation, the brain tissue itself has negligible attenuation to ultrasound. This is the first time that the ultrasonic wave propagation is quantitively mapped inside the brain tissue and can only be achieved by SOPPI.

"Seeing" through an intact skull remains challenging for optical imaging. Although the propagation of light, PA or PT fields inside the skull can't be imaged directly, mapping PA waves across the skull boundary provides insights into understanding acoustic interaction with the skull. We here used an FE to generate the ultrasound and placed it towards the mouse skull (C57BL6J) as shown in the DC image (**Figure 4f**). Upon reaching the skull, the ultrasound experienced a strong reflection due to the high acoustic impedance mismatch between the skull and the water. A maximum intensity mapping of PA signals was performed to visualize the PA field (**Figure 4g**). Despite this, around 21% of ultrasound was able to penetrate the skull and deflect at a small angle, likely due to the skull's uneven surface. A video of the PA waves propagating through the skull was recorded and frames at each 100 ns were shown. (Video S3, **Figure 4h**).

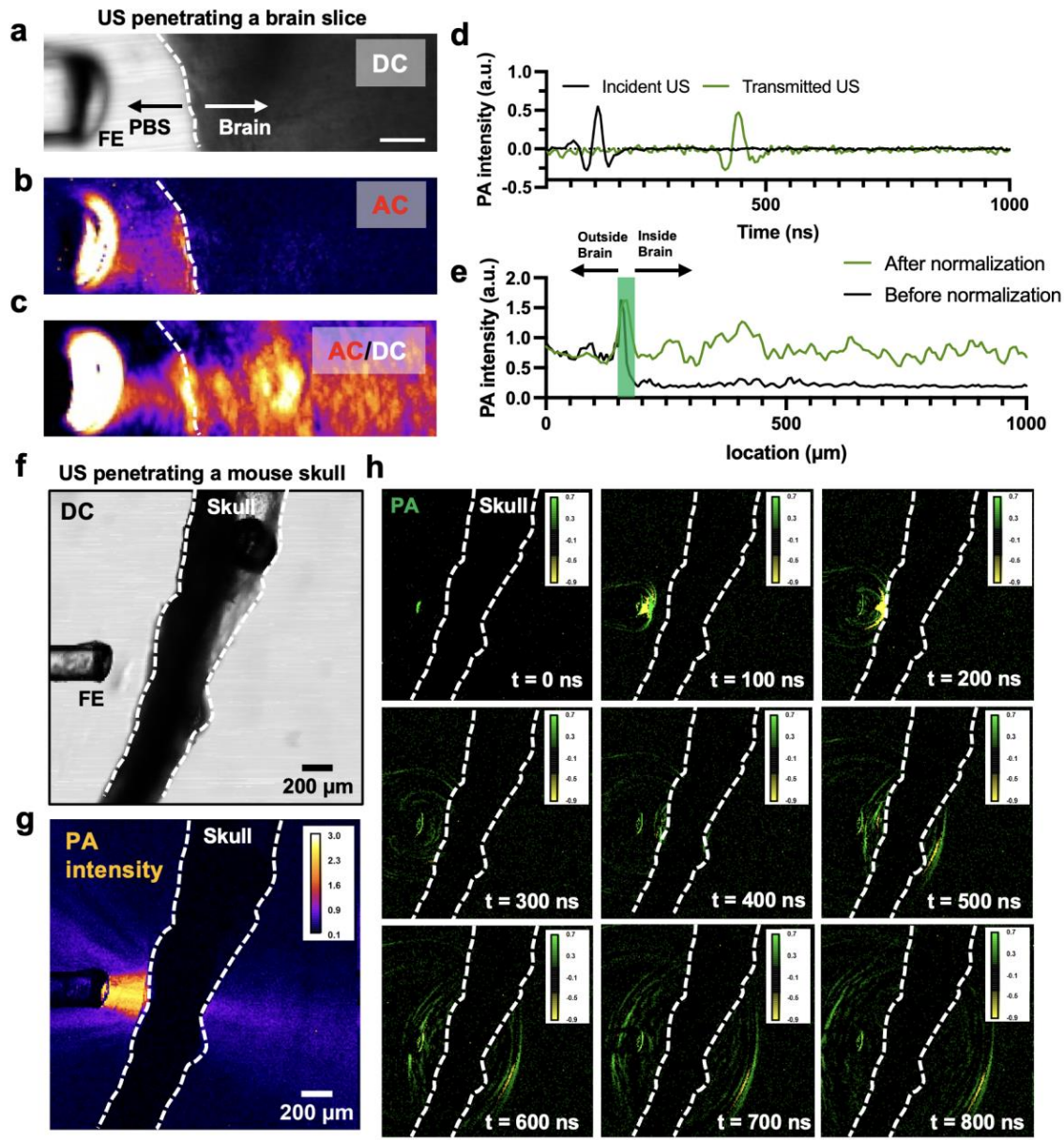

**Figure 4 SOPPI of ultrasound penetrating through mouse brain tissue and skull.** a. transmission DC image of the fiber emitter and a brain slice with 500 μm thickness. Scale bar: 100 μm. b. Spatial distribution of FE-generated PA field in brain slice. c. Spatial distribution of PA signal after normalization with respect to DC signals. d. Representative time trace of the incident / transmitted PA signal. e. PA signal intensity before/after penetrating the brain slice. f. transmission DC image of the fiber emitter and a piece of mouse skull. g. Spatial distribution of the PA field across the skull. h. Nine representative frames of photoacoustic wave penetrating the skull. US: ultrasound.

**SOPPI serves as a programmable virtual transducer array with optical density for photoacoustic computed tomography.**

With its highly sensitive and quantitative detection of PA signals, SOPPI can also reconstruct the absorber distribution within an animal body using a photoacoustic computed tomography (PACT) beamforming algorithm. For demonstration, a zebrafish larva was fixed in agarose gel and illuminated with a 532 nm nanosecond laser as a pump. The scanned probe beam around the zebrafish serves as a programmable virtual transducer array, detecting PA signals emitted by pigments inside its body (**Figure 5a**).

PA signals centered at 30 MHz were captured by SOPPI, leveraging its broad bandwidth detection capability (**Figure 5b**). Using a weighted-delay-and-sum (weighted-DAS) beamforming algorithm, the pigment distribution of the zebrafish was reconstructed, as shown in **Figure 5d.** Details of the algorithm are provided in the Supplementary Information. For reference, a transmission image and the overlaid results are presented in **Figure 5c** and **5e**. In the overlaid images, key anatomical features such as the eyes (E), yolk (Y), swim bladder (B), and dorsal stripe (DS) are clearly identified.

In the reconstructed PACT image, the small pigments on the dorsal stripe are clearly visualized, demonstrating a spatial resolution of 26 μm. This resolution is significantly superior to other PACT systems that use physical transducer arrays. We attribute this improvement to three major factors: broad detection bandwidth, full-angle detection, and high element density.

The broadband detection capability of SOPPI-PACT, as illustrated in **Figure 5b**, features a PA signal centered at 30 MHz with a frequency range extending up to 60 MHz. Higher frequency components enable the recording and reconstruction of finer details in the image. Secondly, full-angle detection plays a crucial role by capturing comprehensive information regardless of the size, location, or geometry of the sample. In a transducer-based PACT system, limited viewing angles pose a significant challenge to imaging quality due to the restricted acceptance angle of linear transducer arrays. Even within their acceptance angle, sensitivity varies with the incident angle of the signals, causing information loss. Addressing this issue typically requires expensive hollow transducer arrays. In contrast, SOPPI employs a probe beam that acts as a virtual transducer and can be adaptively scanned around the sample collecting information emitted from all different angles (**Figure 5f**). This full-angle detection capability significantly reduces image distortion and enables the reconstruction of more detailed information, as shown in **Figure 5f**.

The third advantage of SOPPI-PACT lies in its higher element density, which can surpass traditional transducer arrays by an order of magnitude. The smallest pitch size of commercially available transducer arrays is approximately 100 μm, whereas SOPPI achieves a pitch size of 10 μm. This increase in element density leads to substantial improvements in the signal-to-noise ratio (SNR) of reconstructed images. To simulate the pitch size of traditional transducer arrays, we grouped 10 SOPPI elements into a single equivalent element and reconstructed the zebrafish larva image using the same algorithm. Clear differences can be observed between SOPPI-PACT and simulated traditional PACT (**Figure 5f**). In the traditional PACT image, background noise is significantly higher, and detailed features, such as the pigments on the dorsal stripe, are lost. To quantitatively compare the image quality between SOPPI-PACT and traditional PACT, we calculated the cross- structural similarity index (XSSIM) and cross- peak signal-to-noise ratio (XPSNR) for both methods (**Figure 5g,** detailed definition can be seen in Supplementary Information). The results demonstrate that SOPPI-PACT outperforms traditional PACT in both metrics, further validating the superiority of SOPPI over conventional transducer approaches.

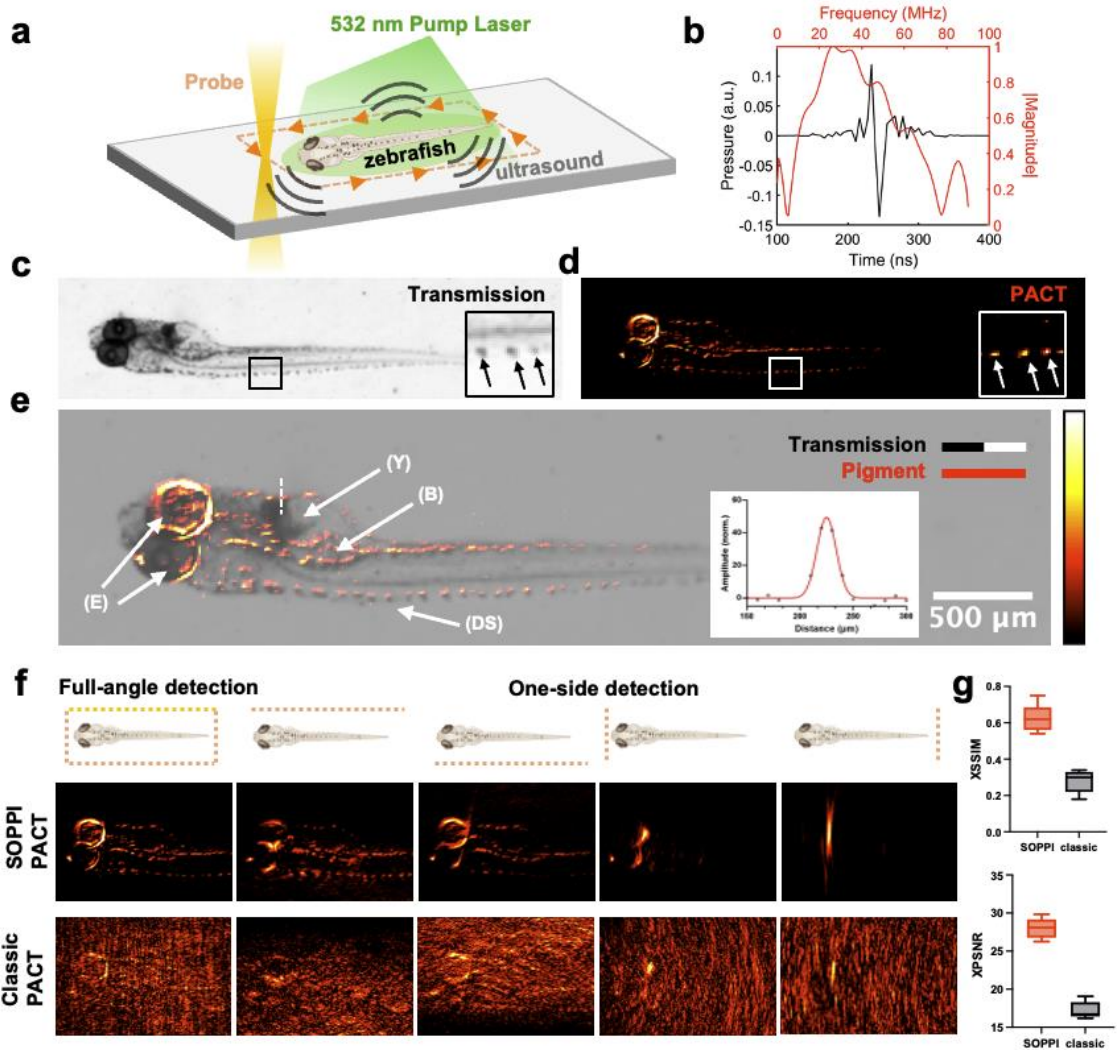

**Figure 5. Photoacoustic computed tomography of a zebrafish using SOPPI.** a. Schematic of a PACT imaging of a zebrafish using SOPPI. The probe beam was scanned around the zebrafish following the yellow dashed line. b. Representative PA signal generated by pigments inside the zebrafish and detected by SOPPI. c. Transmission imaging of a zebrafish. d. Reconstructed mapping of pigments inside the zebrafish using SOPPI-PACT. e. Overlap of transmission and PACT results. f. Comparison of SOPPI-PACT and classic PACT results.

## Discussion

We have developed a scattering/deflection-based spatial offset pump-probe imaging system for mapping fast nonradiative dynamics at optical resolution. PA and PT effects are two major forms of nonradiative relaxation. Compared to traditional transducer-based measurements of the PA field, SOPPI also offers several significant advantages. First, the detection bandwidth of SOPPI is 65 MHz limited by the digitizer, which theoretically could be much improved to even gigahertz with high-frequency photodiodes and digitizers, while traditional piezo-based transducers typically operate within a narrow detection bandwidth. The broad detection bandwidth of SOPPI allows a less biased detection of acoustic waves at different frequencies. Second, transducer-based photoacoustic measurements often suffer from limited view angles due to the varying sensitivity of PA signals at different incident angles. In contrast, the orthogonal probing of SOPPI enables the detection of PA signals from all angles without bias. Lastly, the PA field measured with piezo-based transducers has relatively low spatial resolution constrained by the acoustic wavelength, typically in the range of tens of microns. In contrast, SOPPI intrinsically achieves optical resolution that depends on the numerical aperture of the objective lens and the probe wavelength.

Taking advantage of its high quantitative measurement, optical spatial resolution, and dense scanning capabilities, SOPPI can also function as a virtual ultrasound transducer array for PACT. This SOPPI-PACT system is highly programmable and features an exceptional element density, resulting in significant improvements in SNR compared to traditional transducer array-based PACT. Currently, SOPPI-PACT has been demonstrated on small animals such as zebrafish and holds potential for applications in larger mammalian models, including mice and even humans.

By SOPPI of PA and PT fields generated by an FE, we confirmed that with a 5 ns 1064 nm pulsed laser as the pump, the PT was locally confined in the 200 μm PDMS coating and the heat dissipation outside the FE were undetectable. This result implies that the generated heat has a limited impact on the surrounding environment. Therefore, the non-transgenic neural stimulation with an FE majorly was contributed by the PA rather than the PT effect. Similar to most of pump-probe measurements, SOPPI has limitations to image non-repeatable events. But SOPPI still has a high impact on many applications as mentioned above.

Due to the low repetition rate of the current pump laser, recording a video of 100 μs with a millimeter-level field of view at optical resolution normally takes hours to finish. The total acquisition time could be improved by 100 times by using a pump laser with a kHz repetition rate. A more sensitive detector like an avalanche photodiode could be used to improve the sensitivity of SOPPI. Alternatively, adopted from oblique back-illumination microscopy and oblique photothermal microscopy , the detector could be placed close to the sample on the illumination side and collect the back-propagated transmission photons, which can circumvent the high scattering of turbid samples.

SOPPI opens potentially exciting applications. SOPPI enables the visualization of the frequency and directional properties of the PA field for individual FEs in practical settings, providing valuable guidance for the development and optimization of FEs in real-world applications. Additionally, SOPPI can be employed to study material properties, as PA and PT effects are strongly influenced by mechanical characteristics such as viscosity, Young's modulus, and thermal conductivity. Furthermore, intriguing phenomena, such as the acousto-thermal effect demonstrated in this study, can be explored in greater detail. The tunability of the pump laser further expands its potential by enabling the targeting of different molecules with varying absorption properties.

In summary, SOPPI offers a sensitive way to visualize the fast dynamics of PA and PT waves. The wave evolution, including generation and propagation, in either clear medium or highly scattered tissue samples could be observed. This technique opens new opportunities in helping with the design of PA and PT transducers, visualizing nearfield dynamics, and improving mechanical/thermal-related clinical treatments.

**Materials and Methods**

**FE fabrication**

A multimode fiber (FT200EMT, Thorlabs, Inc., NJ, USA) with 200 µm core diameter was used. The PA coating was composed of candle soot and polydimethylsiloxane (PDMS). Candle soot was chosen as the absorber due to its great absorption coefficient. The multimode optical fiber was exposed to the candle flame for around 3–5 s until the fiber tip was fully coated, with the thickness of the candle soot controlled by the deposition time. To prepare PDMS, the silicone elastomer (Sylgard 184, Dow Corning Corporation, USA) was carefully dispensed into a container to minimize air entrapment, and then mixed with the curing agent in a ratio of 10:1 by weight. A nanoinjector deposited the prepared PDMS onto the tip of the candle-soot-coated fiber and thus formed a layered structure. The position of the fiber and the nanoinjector were both controlled by 3D manipulators (MT3, Thorlabs, Inc., NJ, USA) for precise alignment, and the PDMS coating process was monitored in real-time under a lab-built microscope. The coated fiber was then cured overnight at room temperature.

**Optical fiber tapering**

To control the tapering, a multimode fiber (FT200EMT, Thorlabs, Inc., NJ, USA) was pulled at one end by a traction weight with the other end fixed. The pulling force, determined by the weight of the traction object, was found to be proportional to the square of the tapered end radius.

**SOPPI system**

Spatial-offset pump-probe imaging was performed with a modified microscope. The pump laser was at 1064 nm with a 5 ns pulse duration and 20 Hz repetition rate (OPOLETTE 355 LD, OPOTEK) for PA/PT generation, then coupled into a 200 µm core fiber (M25L01, Thorlabs) using a collimator (F220SMA-1550, Thorlabs). The fiber was connected with other fiber devices, for example, FE for brain slice or tapered fiber for water-generated ultrasound imaging, through a fiber mating sleeve (ADASMA, Thorlabs). A continuous wave 1310 nm laser (1310LD-4-0-0, AeroDIODE Corporation) serves as the probe with a power of 5 mW after the objective. The probe laser was sent to a water dipping objective (UMPLFLN 10XW, Olympus) for illumination and collected by an air objective (MPlanFLN 10x, Olympus). The signal-carrying probe laser was detected by an amplified InGaAs photodiode (PDA05CF2, Thorlabs) with a 1310 nm bandpass filter (FBH1310-12, Thorlabs). The output signal was connected to a 50 Ohm resistor, amplified by a 46 dB amplifier (100 MHz bandwidth, SA230F5, NF corporation), and digitized by a data acquisition card at 180 MSa/s (ATS9462, Alazar Tech), equivalent to 5.6 ns temporal resolution. A translation stage (ProScan III, Prior) was used to scan the fiber that delivered the pump laser and the generated PA/PT fields with a step size tunable from 1 to 20 µm. The data collection was triggered by both the pump pulse and the translation stage. Each pixel corresponds to a single pump pulse.

**Spatial-offset pump-probe imaging of biological tissues**

All experimental procedures have complied with all relevant guidelines and ethical regulations for animal resting and research established and approved by the Institutional animal care and use committee of Boston University (PROTO201800534). Adult C57BL/6 J mice (age 14–16 weeks) were sacrificed and perfused transcardially with phosphate-buffered saline (PBS, 1X, PH 7.4, Thermo Fisher Scientific Inc.) solution and 10% formalin. After fixation, the brain was extracted and fixed in 10% formalin solution for 24 h. The fixed mouse brain was immersed in 1X PBS solution and sliced into coronal sections with a 500 µm thickness using an Oscillating Tissue Slicer (OST-4500, Electron Microscopy Sciences). Brain slices were gently transferred by a brush into 10% formalin solution for another 24 h fixation and fixed on a glass substrate for pump-probe imaging.

**Data analysis**

The pump-probe imaging data was processed through *MATLAB R2023b* and *ImageJ*. The PA and PT traces were created using *MATLAB R2023b* and *Prism9*.

**Acknowledgments**

**Funding:**
National Institutes of Health grant R35 GM136223 (JX C)
National Institutes of Health grant R01 HL125385 (JX C)
National Institutes of Health grant R21 EY036579 (CY)
National Institutes of Health grant R21 EY035437 (CY)

**Author contributions:**
Conceptualization: GC, YY, JX C, CY
Methodology: GC, YY
Investigation: GC, YY, HN, GD, ML, YZ, DL, HZ, HH, ZG
Supervision: JX C, CY
Writing—original draft: GC, YY, JX C, CY
Writing—review & editing: GC, YY, JX C, CY

**Competing interests:** JXC and CY claim COI with Axorus which did not support this work. Other authors claim no COI.

**Data and materials availability:** The data that support the findings of this study are available from the corresponding author upon reasonable request.

# Supplementary Materials for

## Spatial offset pump-probe imaging of nonradiative dynamics at optical resolution

Guo Chen *et al.*

*Corresponding author. Email: jxcheng@bu.edu; cheyang@bu.edu

**This PDF file includes:**

    Supplementary Text
    Figs. S1 to S4

**Other Supplementary Materials for this manuscript include the following:**
    Movies S1 to S3

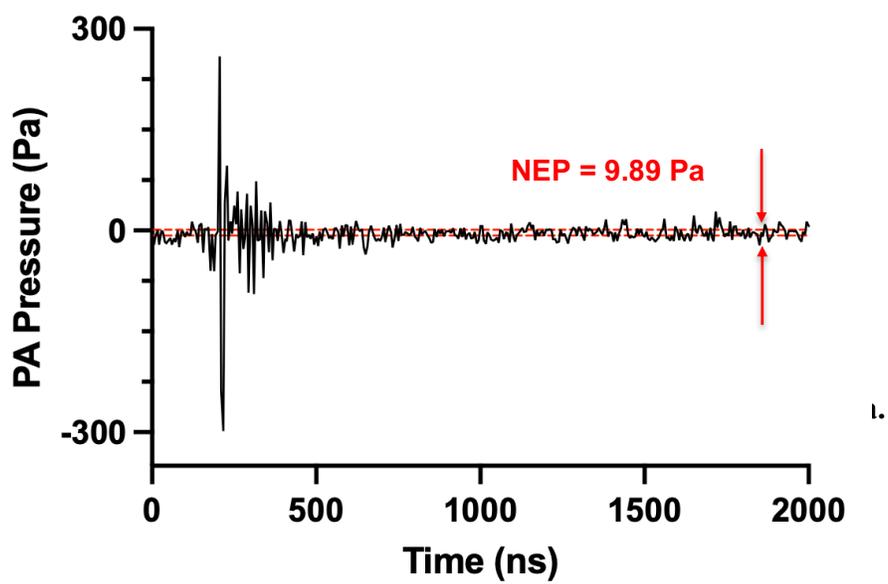

**Figure S1. SOPPI shows a noise equivalent pressure of 9.89 Pa.** 400 times averaged.

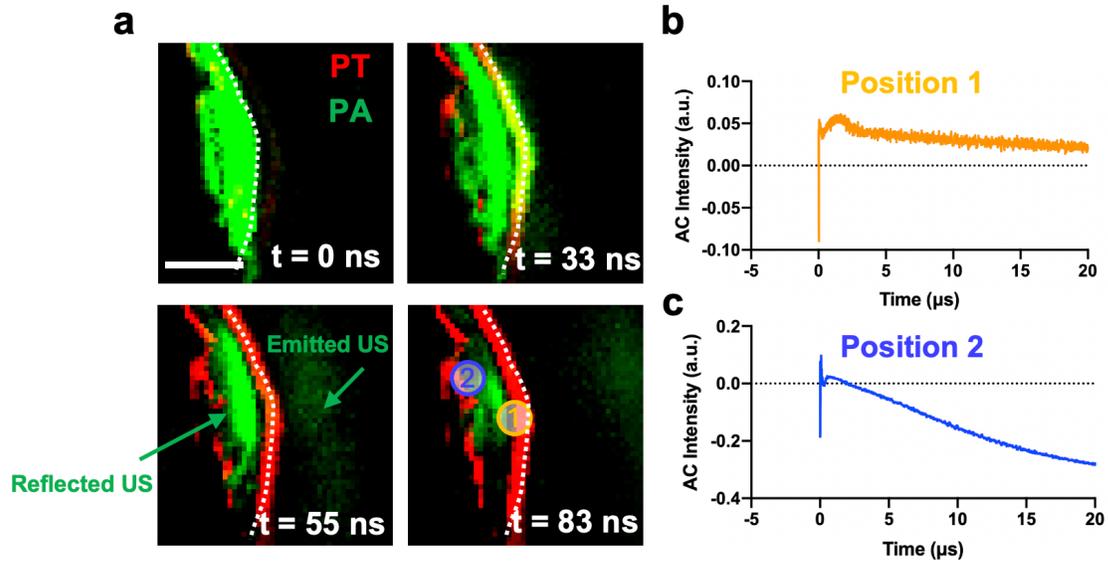

**Figure S2. SOPPI visualizing a near-field acousto-thermal process. a.** Four representative frames of PA and PT field distribution. **b.** signal detected at the boundary of the fiber emitter. **c.** signal detected inside the fiber emitter.

**Calculation of acoustic impedance and reflection ratio on the water / brain boundary**

The acoustic impedance ($Z$) is determined by the density ($\rho$) and the travelling speed of sound ($c$) of the material:
$$Z = \rho \times c$$
In the case of ultrasound penetrating brain slice, the acoustic impedance of water and brain can be calculated as shown below:
$$Z_{water} = 1\ g/ml \times 1480\ m/s = 14.8\ MRy$$
$$Z_{brain\ slice} = 1.08\ g/ml \times 1552\ m/s = 16.7\ MRy$$
With the impedance of both environment and the brain tissue, we can calculate the intensity of ultrasound reflected back:
$$RI^2 = \left(\frac{Z_{water} - Z_{brain}}{Z_{water} + Z_{brain}}\right)^2 = 0.36\%$$
Thus, more than 99% of the acoustic energy will deliver into the brain tissue, which is consistent with our experiment observation.

**Weighted-Delay-and-Sum (W-DAS) SOPPI-PACT reconstruction algorithm**

Delay-and-Sum (DAS) algorithm is widely used in reconstructing photoacoustic computed tomography images. The working principle is shown below in **Figure S3**.

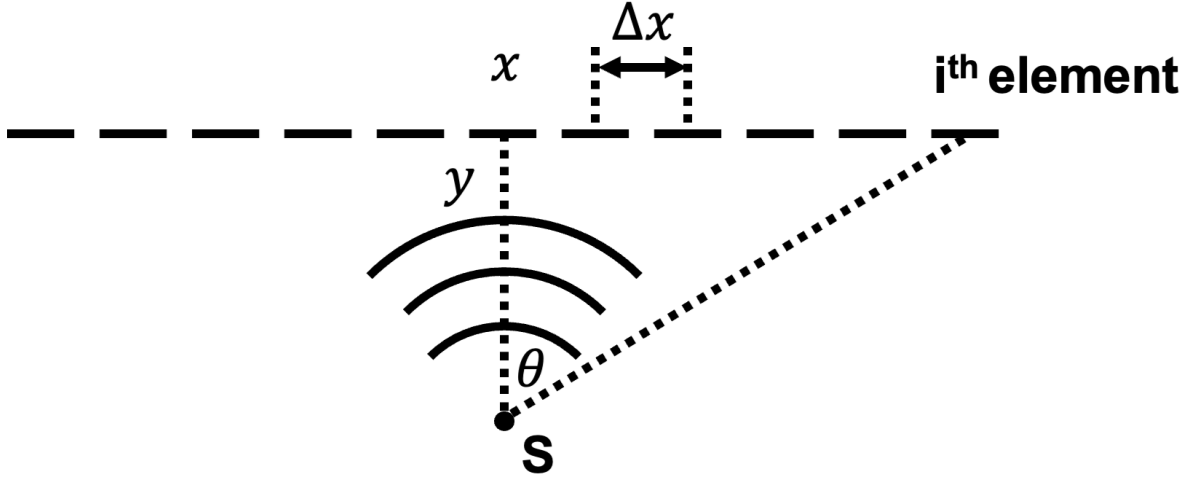

**Figure S3**. The working principle of W-DAS.

Firstly, the DAS algorithm calculates the time delay between the sound source and each transducer elements. We consider each sound source emit a spherical wavefront and travels at a speed of $c$ (1480 m/s for ultrasound in water), and the time delay $\tau$ can be calculated as follows:

$$\tau(x, y, i) = \frac{\sqrt{(x - i \times \Delta x)^2 + y^2}}{c}$$

Then, the beamformed signal $S_{DAS}$ will be:

$$S_{DAS}(x, y) = \sum_{i=1}^{N} \omega(x, y, i) \times S(i, \tau)$$

Where $S$ is the time trace of photoacoustic signal recorded by the i[th] transducer element and $\omega$ is the weight for each transducer element at different location, and $N$ is the total amount of transducer elements. By adding the weight factor, some artifacts like side lobes can be greatly reduced. Here, based on a simple logic that we want to maintain the same element density at any locations and any angles ($\theta$), the weight factor can be calculated as the reciprocal of angular element density (*AED*):

$$AED = \frac{d(y \times tan\theta)}{d\theta} = \frac{y}{cos^2\theta}$$

$$\omega(x, y, i) = \frac{1}{AED} = \frac{cos^2\theta}{y} = \frac{y}{y^2 + (x - i \times \Delta x)^2}$$

Thus, the $S_{DAS}$ will be:

$$S_{DAS}(x, y) = \sum_{i=1}^{N} \frac{y}{y^2 + (x - i \times \Delta x)^2} \times S\left(i, \frac{\sqrt{(x - i \times \Delta x)^2 + y^2}}{c}\right)$$

**Definition of XPSNR and XSSIM**

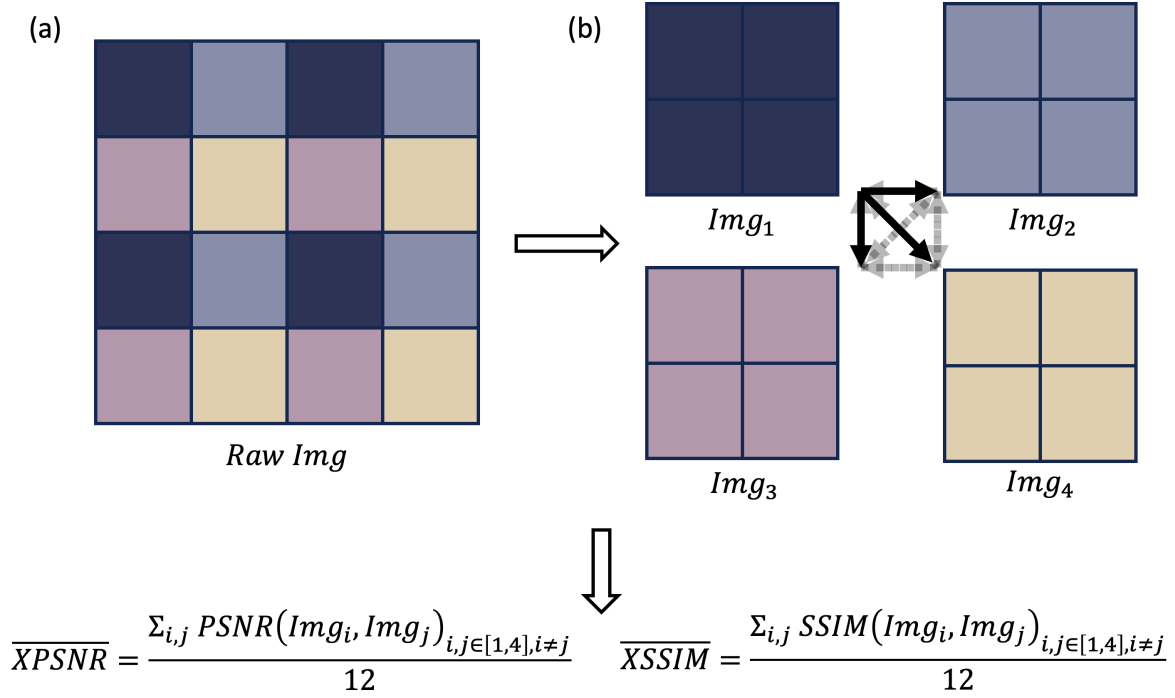

$$\overline{XPSNR} = \frac{\Sigma_{i,j}\, PSNR(Img_i, Img_j)_{i,j\in[1,4], i\neq j}}{12} \qquad \overline{XSSIM} = \frac{\Sigma_{i,j}\, SSIM(Img_i, Img_j)_{i,j\in[1,4], i\neq j}}{12}$$

**Figure S4. Calculation of XPSNR and XSSIM.**

To achieve performance evaluation without ground truth, we develop the Cross Structure Similarity Index (XSSIM) and Cross Peak Signal to Noise Ratio (XPSNR). The original image is split into 4 sub-images, after which SSIM and PSNR are calculated for arbitrary two sub-images pairs and then averaged as shown above.

**Movie S1.**

PA and PT field evolution from a fiber emitter measured by SOPPI

**Movie S2**

PA and PT field evolution from a tapered fiber measured by SOPPI

**Movie S3.**

PA field interacting with a mouse's skull observed by SOPPI